\documentclass{xray_symp}
\usepackage{graphics}
\def\ros{{\sl ROSAT }}
\def\asca{{\sl ASCA }}
\def\G{$\Gamma_{\rm x}$ }
\def\degs{\ifmmode ^{\circ}\else$^{\circ}$\fi}
\def\arcmin{\ifmmode ^{\prime}\else$^{\prime}$\fi}
\def\arcsec{\ifmmode ^{\prime\prime}\else$^{\prime\prime}$\fi}
\def\farcs{\hbox{$.\!\!^{\prime\prime}$}}  
\def\approxlt{\mathrel{\hbox{\rlap{\lower.55ex \hbox {$\sim$}}
        \kern-.3em \raise.4ex \hbox{$<$}}}}
\def\approxgt{\mathrel{\hbox{\rlap{\lower.55ex \hbox {$\sim$}}
        \kern-.3em \raise.4ex \hbox{$>$}}}}
\begin{document}
\title{The \ros view of NGC\,1365: \\
core emission, the highly variable source NGC\,1365--X1, \\ 
and alignments of surrounding X-ray sources}
\author{Stefanie Komossa\inst{1} \and Hartmut Schulz\inst{2}}  
\institute{Max--Planck--Institut f\"ur extraterrestrische Physik,
 Giessenbachstra{\ss}e, 85740 Garching, Germany
\and Astronomisches Institut der Ruhr-Universit\"at Bochum, 44780 Bochum, 
Germany  }
\authorrunning{St. Komossa, H. Schulz}
\titlerunning{The \ros view of NGC\,1365}
\maketitle

\begin {abstract} 

We present a spectral and spatial analysis of the
nuclear and circumnuclear X-ray emission of the 
prominent southern starburst/Seyfert galaxy NGC\,1365.
To describe the X-ray spectrum of the core source, we favour a 
two-component model consisting
of about equally strong contributions from a Raymond-Smith plasma
(with $\sim$cosmic abundances) and a powerlaw.
The origin of both components is discussed, and a 
detailed comparison of the X-ray properties
with multi-wavelength observations of NGC\,1365
and a large sample of type-I AGNs is performed. 

Among the surrounding X-ray sources we focus on an analysis
of the enigmatic source
NGC\,1365-X1, which is one of the most
luminous and most highly variable off-nuclear X-ray sources known so far.

Positions of further X-ray sources in the field of view are derived
and it is briefly pointed out that all HRI sources (except one) are
`aligned' relative to the central source.

\end {abstract}

\section{Introduction}

NGC\,1365 is a  
prominent southern starburst/Seyfert galaxy.
Its nuclear and disk emission-line gas has been
investigated in numerous optical studies (e.g., Burbidge \& Burbidge 1960,
J\"ors\"ater et al. 1984, Schulz et al. 1994, Lindblad et al. 1996, Roy \& Walsh 1997).  
The presence of an AGN was first suggested by Veron et al.\ (1980) who
found  broad emission-line H$\alpha$ indicative of a Seyfert-1.5 galaxy.
Surprisingly, just in the nucleus identified
by Edmunds \& Pagel (1982), Seyfert-typical narrow-line emission
line ratios are missing. 
In the circumnuclear emission-line region, HII region-like line
ratios
are common indicating widespread circumnuclear star formation.

Summarizing, optical studies suggest that the central region
of NGC\,1365 consists of an AGN of apparent low luminosity
surrounded by a region of enhanced
star formation.
However, the
relationship between the stellar and nonthermal activity
and the geometry of the nucleus need further scrutiny.

X-rays are an important probe of the central activity.
Here, we present an analysis of \ros (Tr\"umper 1983) PSPC and HRI X-ray 
observations 
of the core of NGC\,1365 and an extraordinary
off-nuclear X-ray source which we term NGC\,1365--X1 (for details see 
Komossa \& Schulz 1998). 

\section{The core source}

\subsection {X-ray spectrum}

We studied several spectral models to explain the soft X-ray spectrum.
Successful fits can be obtained by either (i) a single 
\underline{R}aymond-\underline{S}mith model (RS)
but with strongly depleted abundances of $\sim 0.1 \times$ cosmic,  
(ii) two component models consisting of a double RS or an RS plus 
\underline{p}ower\underline{l}aw (PL),
or (iii) a warm `absorber' seen in emission and reflection (`warm 
scatterer').   

Since the low inferred abundances of the single RS model 
are inconsistent with abundance determinations
on the basis of optical data (e.g., Alloin et al. 1981,
Roy \& Walsh 1988, 1997)
 we favour the two component models. 
This is in line with recently reported ASCA observations by Iyomoto et 
al. (1997),
who detect a PL component, and particularly, a strong Fe line in the X-ray
spectrum of NGC\,1365.  

We interpret the PL component to arise from the active nucleus, the RS 
component
to be related to the starburst. 

\subsection {Starburst component}

A 100\degs~wide [OIII] enhanced region 
to the SE of the nucleus
was kinematically modelled as an outflow cone
by Hjelm \& Lindblad (1996). This could be a Seyfert outflow, driven
by a wind from the active core, or a starburst outflow, driven by
a series of supernova explosions. Assuming that the outflow region
 evolved from a wind-driven
supershell of swept-up ISM we can 
estimate the starburst contribution to $L_{\rm x}$
(generated in the shell and the
bubble interior to it).

As will be shown below the optically detected AGN could only
provide
a few percent of the IR emission.
Assuming that star formation takes care of the rest
we take the IRAS $L_{\rm IR}= 2.4\,10^{44}$ erg/s as an estimate
for the bolometric luminosity of the starburst for which
models from Gehrz et al.\ (1983)  predict a SN rate between 0.01
and 1 yr$^{-1}$ 
and $L({\rm H}\alpha)$ in the range $7\,10^{41} - 7\,10^{42}$ erg/s
of which Kristen et al. (1997) see only $6\,10^{40}$ erg/s in the central
region.
Hence, 90\% to 99\% of the H$\alpha$ emitters ionized by young stars are
hidden.
Could this hidden burst supply the X-rays in a wind driven shell?

Using the analytical models of MacLow \& McCray (1988) we find that 
this crude scenario can account
for the observed thermal X-ray luminosity
if more than 90\% of the ionized
gas emitting optical lines in the starburst region is obscured.

\begin{figure} 
\vspace{0.3cm}
\resizebox{\hsize}{!}{\includegraphics{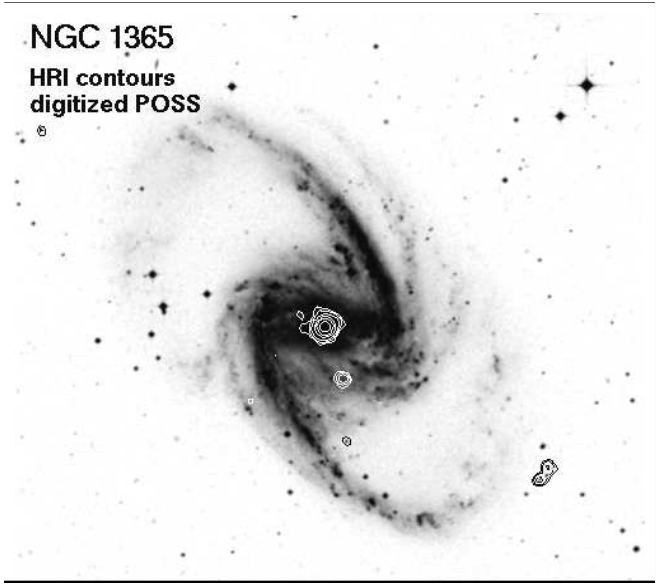}}
\caption[over]{Overlay of the HRI X-ray contours on an optical image
of NGC\,1365 from the digitized POSS. The source south-west to the core
source, located on one of the subordinate spiral arms, is NGC\,1365-X1. 
The source further to the south-west, `companion 2' of Turner et al. 
(1993),
is resolved in two sources. 
}
\label{over}
\end{figure}

\subsection{AGN component}

We now compare the data with the sample of hard X-ray detected AGN
from Piccinotti et al.\ (1982). 
For this sample, Ward et al. (1988)
showed that the luminosities
of hard X-rays,
H$\alpha$ and  mid-IR radiation
are well correlated.
We represent their results by the relations
\begin{eqnarray}
\log L_{\rm 2-10 keV} & = & 0.947 \log L({\rm H}\alpha) + 3.447~~, 
\nonumber
\\
\log L_{\rm 2-10 keV} & = & 1.326 \log L(25-60\mu {\rm m}) - 15.283~~.
\end{eqnarray}

For NGC\,1365 
it turns out that the observed values 
($\log L_{\rm 2-10 keV} = 41.08$, $\log L({\rm H}\alpha_{\rm broad}) = 
40.82$,
$\log L(25-60\mu {\rm m}) = 44.87$),
do not fit the
relations
of the
Piccinotti sample for which we assume that it defines
`pure' AGN properties.
E.g., if the hard X-rays come from the AGN in
NGC\,1365
we would expect $\log L(25-60\mu {\rm m}) = 42.5$ which is only 0.4\%
of the
observed 44.87. The remaining IR could either be attributed to star
formation
or the source for the hard 2-10 keV photons is for a large part obscured
which would, however,
require column densities exceeding $10^{24}$ cm$^{-2}$.

Using the first Piccinotti-sample relation of Eq.\ 1,
the X-rays predict $\log L({\rm H}\alpha)=39.74$, only 8.3\% of the
observed
40.82. 
A possible explanation would be some arrangement of obscuring material in 
front of
the X-ray source or/and the BLR, possibly combined with
scattering elsewhere. 

Taking H$\alpha$ as representative
for the AGN, $\log L_{\rm 2-10 keV} = 42.10$, i.e.\ ten times the
observed hard X-ray
luminosity 
 would be predicted. This type of solution would fit to our
warm scatterer model which could explain both the
hard PSPC
(and {\sl ASCA}) PL-like component and the high equivalent width of the
FeK
complex (like in NGC\,6240, see Komossa et al. 1998).

\subsection{Place within the unified model}

NGC\,1365 (with its strong BLR component and weak X-ray emission)
does not fulfill expectations of the simplest version of the unified model
in which only the torus blocks the light and detection of a BLR would
imply an unobscured view of the X-ray source as well.
Complicated models appear to be necessary to let this AGN be an
{\em intrinsically normal} broad-line object.

\section{The luminous and highly variable source \mbox{NGC\,1365--X1}}

The HRI data clearly resolve the nuclear emission from that of the 
off-nuclear source RX\,J0333-36  = NGC\,1365--X1, and locate the latter
on one of the spiral arms (Fig. \ref{over}). 

\begin{figure} 
\resizebox{\hsize}{!}{\includegraphics{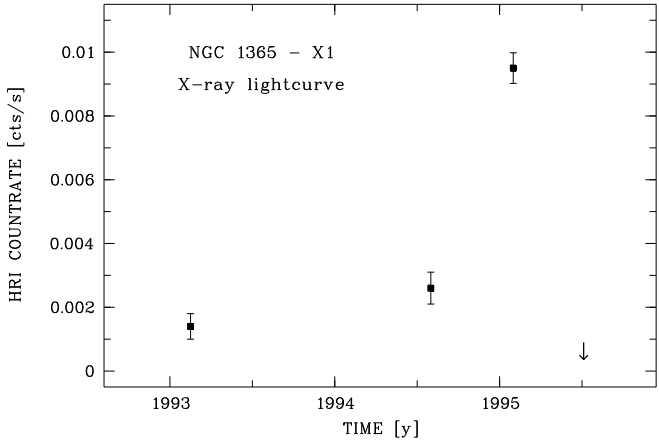}} 
 \caption[light_offn]{Long-term X-ray lightcurve of NGC\,1365--X1
(all count\-rates were converted to \ros HRI counts). The source is
strongly variable on the timescale of months.
}
 \label{light_offn}
\end{figure}

The source turns out to be highly variable on the
timescale of months. So far, the high-state was reached during the
\asca observation by Iyomoto et al. (1997). We detect a drop in 
luminosity 
by a factor of more then ten in the second \ros HRI observation taken
a few months later. The complete X-ray lightcurve is given in Fig. 
\ref{light_offn}.

Intrinsic to NGC\,1356, the source is exceptionally luminous, with 
$L_{\rm 2-10 keV} \approxgt 4\,10^{40}$ erg/s in the high-state.
Its temporal variability excludes an interpretation in terms of several
spatially unresolved weak sources.
Although a supernova in dense medium would be an efficient way to reach
high X-ray luminosities (e.g., Shull 1980, Vogler et al. 1997), 
the huge variability on the timescale
of months we detect seems to favour an interpretation in terms
of accretion onto a compact object.

At present, the most likely interpretation seems to be an ultra-powerful
X-ray binary, with either a highly super-eddington low-mass black hole
or a very massive black hole.
In case the accretion is not super-eddington, a rather
high-mass black hole of $\sim$100--200 M$_{\odot}$ is inferred
which may pose a
challenge
for stellar evolution models.

\begin{table}[h]             
\caption{Summary of the properties of NGC\,1365--X1.} 
\begin{center}
     \label{fitres2}
      \begin{tabular}{cccc}
      \hline
      \noalign{\smallskip}
\multicolumn{4}{l}{HRI position (J\,2000)}  \\
\multicolumn{4}{l}{~~~ $\alpha = 3^h 33^m 34.5^s,~~ \delta = -36\degs
9\arcmin 38\farcs$0} \\
\multicolumn{4}{l}{spectral properties} \\
\multicolumn{4}{l}{~~~ \G $\simeq -1.5$ for $N_{\rm Gal}$, $L_{\rm x} 
\simeq
2.4\,10^{39}$erg/s (PSPC low-state) } \\
\multicolumn{4}{l}{variability}\\
\multicolumn{4}{l}{~~~ amplitude factor $\approxgt$ 10, within
 $t$ $\approxlt$ 6 months}\\
      \noalign{\smallskip}
      \hline
  \vspace{-0.4cm}
 \end{tabular}
\end{center}
\end{table}
%

\section{Further X-ray sources in the field of view}

In total, nine and eight X-ray sources with $> 3\sigma$ are detected 
within the f.o.v
of the first and second HRI observation (dubbed HRI-1 and HRI-2), 
respectively, and 
a few additional weak sources with the PSPC.  
Several sources are variable. 

The position of the brightest X-ray source in the f.o.v. is in excellent 
agreement
with that of the BL Lac object MS\,03313-36, enabling a rather safe 
identification.
Another source coincides with FCC\,129 (see Fig. 3), a low-surface brightness galaxy 
that may belong to the Fornax cluster.
Most other sources are presently unidentified.
In particular, none coincides with one of the optical supernovae detected 
in NGC\,1365 (we note that `companion 4' of Turner et al. (1993, their Tab. 2A,B) 
is within 16-20\arcsec of SN\,1957C; Fig. 3). 

\begin{figure}[h]      
\resizebox{\hsize}{!}{\includegraphics{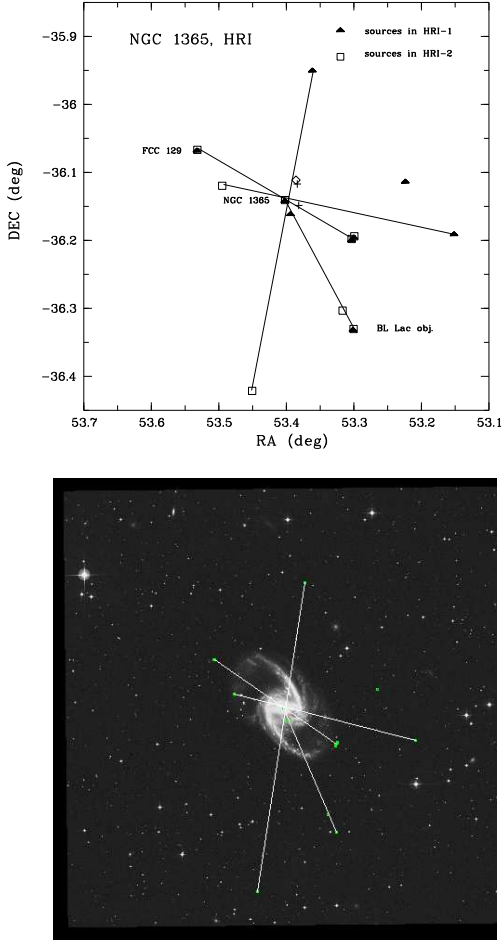}} 
      \vspace{-0.6cm}
 \caption[arp1]{Positions of X-ray sources in the vicinity of the nucleus 
of NGC\,1365.
Upper panel: Sources detected with $>3\sigma$ in the first (filled 
triangles)
and second (open squares) HRI observation. 
In addition, the positions of the two optical supernovae in NGC\,1365,
SN 1957C (upper cross) and SN 1983V (lower cross) and 
Turner et al.'s `companion 4' (lozenge) are marked.
Lower panel: HRI source positions 
on top of an
optical image of NGC\,1365 from the digitized POSS. }
 \label{arp1}
\end{figure}

In Fig. \ref{arp1} we plot the positions of all HRI sources.
Among the nine sources detected in HRI-1, five spatial coincidences are 
found in
HRI-2, the remaining three detected with HRI-2 were below the threshold 
in HRI-1.
In total, twelve sources were detected by the HRI observations.
The nuclear source was detected in both observations with comparable 
count rates.
Taking the nucleus as the origin (0,0) of a rectangular coordinate system
(we are dealing here with a field of $\pm 0\fdg2$ so that we can neglect
effects of spherical trigonometry in a good approximation) with the
$x$-axis from west to east (ascending RA) and the $y$-axis from south to
north (ascending declination), eleven
extranuclear HRI sources are left. 

Sources with nearly the same polar angle $\phi$ in this coordinate system 
are considered
`to be aligned'. Those `on the other side of the nucleus' have nearly the 
same
polar angle `mod 180\degr'. We found the following `alignments' with the
nucleus (i.e. at least two sources are aligned with the nuclear source) 
or `groupings within
a few degrees' (polar angles and the corresponding `mod 180\degr' are 
given):
Group A1: $12\fdg7$ and $191\fdg1 \cor 11\fdg1$;
group A2: $29\fdg6$, $207\fdg6 \cor 27\fdg6$ and $210\fdg2 \cor 30\fdg2$;
group A3: $241\fdg9$ and $242\fdg4$ and $246\fdg7$;
group A4: $279\fdg9 \cor 99\fdg9$ and $102\fdg0$;
one object with p.a. $171\fdg0$ is remaining with no `aligned' 
counterpart 
above the threshold of the HRI observations.
The maximal observed angular spreads in each group are $1\fdg6$, 
$2\fdg6$, $4\fdg8$ and 
$2\fdg1$, respectively.

A few tests were commenced
to find out whether these alignments suggest a relationship to
the nucleus of NGC\,1365. To this end, employing a random-number generator
we produced 
eleven positions around the center inside a flat square, with uniformly
distributed components $x$ and $y$.
Then the polar angles $\phi$ modulo 180\degr~were determined and we 
looked for groupings among them.

The question arises how to define a situation similar to the observed one.
A crude feature of the observations is the occurrence of six angular
differences $\Delta \phi < 5\degr$ (several are significantly smaller!). 
In a first step, we checked
which fraction of the simulations fullfil this criterion.
At large numbers ($>1000$), we found 1.2\% with six $\Delta \phi < 
5\degr$; in 32\% of the
cases (the peak of the histogram) there are only three occurrences.
We looked at a few particular `successes' and found that none of them 
reaches the
tightness of the observed alignments. 

Hence, to interpret the observations by chance alignments requires an 
extremely
rare event. To proceed further in these matters, optical identifications 
and
subsequent spectroscopy, if possible, are necessary and a deliberate 
search
for similar situations around other galaxies would be required. 
For a study of alignments of {\em bright} X-ray sources near galaxies see
Arp (e.g., 1997 and references therein) and Burbidge (1998, these proceedings).

\section {Summarizing conclusions} 

We have presented an X-ray analysis of the
composite starburst/Seyfert galaxy NGC\,1365.
Excellent fits of the \ros PSPC spectrum
are obtained by combining a soft thermal component with a hard
powerlaw.
The hard component may be either seen directly
or can be explained by
scattering of the AGN powerlaw at circumnuclear warm high-column-density
gas.

A compilation of the multi-wavelength properties of NGC\,1365
and comparison with hard X-ray
selected AGNs shows that the hard component of NGC\,1365 is too
faint compared to its broad Balmer line components challenging
simple unified models.

According to analytical estimates, supernova driven outflow can fully
account for the X-ray luminosity in the Raymond-Smith component
if the observed IR emission is mainly provided by the central starburst.

We found that ten of the eleven HRI-detected sources fall into
four `alignment groups' which are in polar angle mod 180\degr (the slope 
angle of the line 
connecting the source with the nucleus) each closer together than 5\degr.
Further investigations are required to check whether this is one of the
rare possible chance occurrences. 

With the \ros HRI data, we have precisely located the extraordinary
southwest X-ray source
NGC\,1365--X1 which falls on one of the spiral arms. 
The source is found to be highly variable
(a factor $\approxgt$ 10) on the timescale of months.
Intrinsic to NGC\,1365, its huge luminosity makes it exceptional among 
stellar
X-ray sources. At present, the most likely interpretation seems to be
an ultra-powerful X-ray binary.

\begin{acknowledgements}
St.K. acknowledges support from the Verbundforschung under grant No. 50\,OR\,93065.
It is a pleasure to thank Per Olof Lindblad for helpful comments and suggestions,
and Andreas Vogler for providing the software
to plot the overlay contours in Fig. 1.
The \ros project is supported by the German Bundes\-mini\-ste\-rium
f\"ur Bildung, Wissenschaft, Forschung und Technologie
(BMBF/DLR) and the Max-Planck-Society.
\end{acknowledgements}

\end{document}